\documentclass [12pt]{article}
\usepackage {graphicx}
\usepackage{epsfig,amsmath,latexsym,amssymb}
\usepackage{bm}
\usepackage{booktabs}
\usepackage{multirow}
\renewcommand{\baselinestretch}{1.7}

\begin{document}

\title{Bayesian Model Choice of Grouped t-copula}

\author{Xiaolin Luo$^{1\ast}$ \quad Pavel V.~Shevchenko$^{2}$}

\date{\footnotesize{This version 2 March 2011 }}

\maketitle

\begin{center}
\footnotesize { \textit{$^{1}$ CSIRO Mathematics, Informatics and Statistics, Australia; e-mail: Xiaolin.Luo@csiro.au \\
$^{2}$ CSIRO Mathematics, Informatics and Statistics, Australia;
e-mail: Pavel.Shevchenko@csiro.au  \\
$^*$ Corresponding author} }
\end{center}

\begin{abstract}
\noindent One of the most popular copulas for modeling dependence
structures is $t$-copula. Recently the grouped $t$-copula was
generalized to allow each group to have one member only, so that
\textit{a priori} grouping is not required and the dependence
modeling is more flexible. This paper describes a Markov chain Monte
Carlo (MCMC) method under the Bayesian inference framework for
estimating and choosing $t$-copula models. Using
 historical data of foreign exchange (FX) rates  as a case study, we found that Bayesian
model choice criteria overwhelmingly favor the generalized
$t$-copula.  In addition, all the criteria also agree on the second
most likely model and these inferences are all consistent with
classical likelihood ratio tests. Finally, we demonstrate the impact
of model choice on the conditional Value-at-Risk for portfolios of
six major FX rates.

\vspace{1cm} \noindent \textbf{Key words:} grouped $t-$copula,
  dependence modeling, Bayesian model choice, Markov chain Monte Carlo,
 foreign exchange.
\end{abstract}

\pagebreak

\section{Introduction}
\label{sec:introductionords}

Copula functions have become popular and flexible tools in modeling
multivariate dependence among financial risk factors. In practice,
one of the most popular copulas in modeling multivariate financial
data is perhaps the $t$-copula implied by the multivariate
$t$-distribution (hereafter referred to as {\it standard t-copula});
see \nocite{EmbMS01} Embrechts \textit{et al} (2001),
\nocite{FangFK02} Fang \textit{et al} (2002), and \nocite{DemM05}
Demarta and McNeil (2005). This is due to its simplicity in terms of
simulation and calibration, combined with its ability to model tail
dependence which is often observed in financial returns data. Papers
by \nocite{MasNZ03} Mashal \textit{et al} (2003) and
\nocite{BreDE03} Breymann \textit{et al} (2003) have demonstrated
that the empirical fit of the standard $t$-copula is superior in
most cases when compared to the Gaussian copula. However, the
standard $t$-copula is often criticized due to the restriction of
having only one parameter for the degrees of freedom (dof), which
may limit its ability to model tail dependence in  multivariate
case. To overcome this problem, \nocite{DauGL03} Daul \textit{et al}
(2003) proposed the use of the {\it grouped t-copula}, where risks
are grouped into classes and each class has its own standard
$t$-copula with a specific dof. This, however, requires an \textit{a
priori} choice of classes. It is not always obvious how the risk
factors should be divided into sub-groups. An adequate choice of
grouping configurations requires substantial additional effort if
there is no natural grouping, for example, by sector or class of
asset.

Recently, the grouped $t$-copula was generalized to a new $t$-copula
with multiple dof parameters (hereafter referred to as {\it
generalized t-copula}); see \nocite{LuoS09} Luo and Shevchenko
(2010) and \nocite{Venter07} Venter et al (2007). This copula can be
viewed as a grouped $t$-copula with each group having only one
member. It has the advantages of a grouped $t$-copula with flexible
modeling of multivariate dependences, yet at the same time it
overcomes the difficulties with \textit{a priori }choice of groups.
For convenience, denote the new copula as $\tilde {t}_{\bm \nu
}$-copula, where ${\bm \nu }=(\nu_1,...,\nu_n)$ denotes the vector
of dof parameters and $n$ is the number of dimensions. Luo and
Shevchenko (2010) demonstrated that some characteristics of this new
copula in the bivariate case are quite different from those of the
standard $t$-copula. For example, the copula is not exchangeable if
$\nu_1 \neq \nu_2$ and tail dependence implied by the $\tilde
{t}_{\bm \nu } $-copula depends on both dof parameters. The
difference between $\tilde {t}_{\rm {\bm \nu }} $- and standard
$t$-copulas, in terms of impact on Value-at-Risk (VaR) and
conditional Value-at-Risk (CVaR) of the portfolio, can be
significant as demonstrated by simulation experiments for the
bivariate case. This difference is even much larger than the
difference between Gaussian copula and the standard $t$-copula. In
examples of maximum likelihood  fitting to USD/AUD and USD/JPY daily
return data, standard $t$-copula was statistically rejected by a
formal Likelihood Ratio test in favour of the $\tilde {t}_{\bm \nu }
$ copula (i.e. dof parameters in the $\tilde {t}_{\bm \nu  }
$-copula were statistically different).

This paper presents a Bayesian model selection study on the
$t$-copula models in the multivariate case. We demonstrate how to
perform Bayesian inference using Markov chain Monte Carlo (MCMC)
simulations to estimate parameters and make decisions on model
choice. From a Bayesian point of view, model parameters are random
variables whose distribution can be inferred by combining the prior
density with the likelihood of observed data. The complete posterior
distribution of the parameters resulting from Bayesian MCMC allows
further analysis such as model selection and parameter uncertainty
quantification. Specifically, we solve a variable selection problem
in the same vein as discussed in Cairns (2000). \nocite{Cairns00}
Increasingly, Bayesian MCMC finds new applications in quantitative
financial risk modeling. Recent examples  are found in
\nocite{PeShWu09} \nocite{PeShWu10} Peters \textit{et al}. (2009,
2010) for  insurance, \nocite{Shevchenko10} Shevchenko (2010) for
operational risk and \nocite{LuoSLGD} Luo and Shevchenko (2010) for
credit risk.

As a case study, we consider the application of modeling  dependence
among six major foreign exchange (FX) rates (AUD, CAD, CHF, EUR, GBP
and JPY, against USD) using $t$-copulas.  Following common practice
(see e.g. \nocite{McFrEm05} McNeil \textit{et al} 2005),  we use the
GARCH(1,1) model to standardize the log-returns of the exchange
rates marginally. Then the GARCH filtered residuals of the six major
FX rates are modeled by a $t$-copula.   In this study we consider
altogether 33 competing $t$-copula models:
 the standard $t$-copula, 31 grouped $t$-copulas and the
generalized $t$-copula (i.e. $\tilde {t}_{\bm \nu } $-copula). The
31 grouped $t$-copulas are a complete set of all possible
combinations of two groups from six FXs (see Table 1 for all
possible 2-group configurations for the six FX majors).

 We present
procedures and results of MCMC simulation for $t$-copula models
under the Bayesian framework. Also, we  demonstrate using  Bayesian
model inference and  actual data, that the generalized $t$-copula
($\tilde {t}_{\bm \nu } $-copula) is convincingly the model of
choice for modeling dependence between six FX majors, among
considered 33 $t$-copula models. Even compared with the best grouped
$t$-copula chosen from  31 possible two-group configurations, the
$\tilde {t}_{\bm \nu } $-copula is overwhelmingly favoured by the
Bayesian factors obtained from the MCMC posterior distribution. We
demonstrate that the joint calibration of grouped $t$-copula can be
done very efficiently by applying MCMC. Using model parameters
estimated from  MCMC, we also demonstrate the impact of model choice
on CVaR of two portfolios of six FX majors.

The organisation of this paper is as follows. Section 2 introduces
the various $t$-copula models and notations. Then it describes the
 GARCH model filtering for the six FX
majors, and calibration of the $t$-copula models using the maximum
likelihood method. Section 3 discusses the Bayesian inference
formulation, the MCMC simulation algorithm, the reciprocal
importance sampling estimator and the deviance information criterion
 for model selection. Direct computing of the posterior model probability is also discussed
in Section 3. Section 4 presents MCMC results and the corresponding
Bayesian model selection, in comparison with the traditional maximum
likelihood results and Likelihood Ratio tests.  Examples of
portfolio CVaR calculation using selected models and calibrated
parameters are provided in Section 5, demonstrating the impact of
model choice on risk quantification. Concluding remarks are given in
the final section.

\section{Model, data and  maximum likelihood calibration }

It is well known from  Sklar's theorem (see \nocite{Sklar59} Sklar
1959 and \nocite{Joe97}Joe 1997) that any joint distribution
function $F$ with continuous (strictly increasing) margins $F_1 ,F_2
,\dots,F_n $ has a unique copula
\begin{equation}
C({\rm {\bf u}}) = F(F_1^{ - 1} (u_1 ),F_2^{ - 1} (u_2
),\ldots,F_n^{ - 1} (u_n )).
\end{equation}
The $t$-copulas are most easily described and understood by a
stochastic representation, as defined below.

\subsection{t-copula models }

We introduce notation and definitions as follows:
\begin{itemize}
\item ${\rm {\bf Z}} = (Z_1 ,\dots,Z_n {)}'$ is a random vector from the
multivariate normal distribution $\Phi _{\rm {\bf \Sigma }} ({\rm
{\bf z}})$ with zero mean vector, unit variances and correlation
matrix ${\rm {\bf \Sigma }}$.

\item ${\rm {\bf U}} = (U_1,\dots,U_n {)}'$ is defined on $[0,1]^n$
domain.

\item $V$ is a random variable from the uniform (0,1) distribution
independent of ${\rm {\bf Z}}$.

\item $W = G_\nu ^{ - 1} (V),$ where $G_\nu (\cdot)$ is the distribution
function of $\sqrt {\nu / S} $ with $S$ distributed from the
chi-square distribution with $\nu $ dof, i.e. $W$ and ${\rm {\bf
Z}}$ are independent.

\item $t_\nu (\cdot)$ is the standard univariate $t$-distribution and
$t_\nu ^{ - 1} (\cdot)$ is its inverse.
\end{itemize}

Then we have the following representations.

\vspace{0.2cm} \noindent \textbf{Standard
}\textbf{\textit{t}}-\textbf{copula}

\noindent The random vector
\begin{equation}
{\rm {\bf X}} = W\times {\rm {\bf Z}}
\end{equation}
\noindent is distributed from a multivariate $t$-distribution and
random vector
\begin{equation}
{\rm {\bf U}} = (t_\nu (X_1 ),\dots,t_\nu (X_n ){)}'
\end{equation}
\noindent is distributed from the standard $t$-copula.

\vspace{0.2cm} \noindent\textbf{Grouped}
\textbf{\textit{t}}-\textbf{copula}

\noindent Partition $\{1,2,\dots,n\}$ into $m$ non-overlapping
sub-groups of sizes $n_1 ,\dots, n_m $. Then the copula of the
random vector
\begin{equation}
{\rm {\bf X}} = (W_1 Z_1 ,\dots,W_1 Z_{n_1 } ,W_2 Z_{n_1 + 1}
,\dots,W_2 Z_{n_1 + n_2 } ,\dots,W_m Z_n {)}',
\end{equation}
\noindent where $W_k = G_{\nu _k }^{ - 1} (V)$, $k = 1,\dots,m$, is
the grouped $t$-copula. That is,
\begin{equation}
{\rm {\bf U}} = (t_{\nu _1 } (X_1 ),\dots,t_{\nu _1 } (X_{n_1 }
),t_{\nu _2 } (X_{n_1 + 1} ),\dots,t_{\nu _2 } (X_{n_1 + n_2 }
),\dots,t_{\nu _m } (X_n ){)}'
\end{equation}
\noindent is a random vector from the grouped $t$-copula. Here, the
copula for each group is a standard $t$-copula with its own dof
parameter (i.e. $\nu_k$ is dof parameter of the standard $t$-copula
for the $k$-th group).

\vspace{0.2cm} \noindent \textbf{Generalized
\textit{t}}-\textbf{copula with multiple dof (}$\tilde {t}_{\bm \nu
} $-copula)

\noindent Consider the grouped $t$-copula where each group has a
single member. In this case the copula of the random vector
\begin{equation}
{\rm {\bf X}} = (W_1 Z_1 ,\,W_2 Z_2 ,\dots,W_n Z_n {)}'
\end{equation}
\noindent is said to have a $t$-copula with multiple dof parameters
${\bm \nu}=(\nu_1,\dots , \nu_n)$, which we denote as $\tilde
{t}_{\rm {\bm \nu }} $-copula. That is,
\begin{equation}
{\rm {\bf U}} = (t_{\nu _1 } (X_1 ),t_{\nu _2 } (X_2),\ldots, t_{\nu
_n } (X_n ){)}'
\end{equation}
\noindent is a random vector distributed according to  $\tilde
{t}_{\rm {\bm \nu }} $-copula. Note, all $W_i $ are perfectly
dependent.

\vspace{0.2cm}

 \noindent {\bf Remark:} Given the above
stochastic representation, simulation of the $\tilde {t}_{\bm \nu }
$ copula is straightforward. In the case of standard $t$-copula $\nu
_1 = \cdots = \nu _n = \nu $ and in the case of grouped $t$-copula
the corresponding subsets have the same dof parameter. Note that,
the standard $t$-copula and grouped $t$-copula are special cases of
$\tilde {t}_{\bm \nu }$-copula.

From the stochastic representation (6-7), it is easy to show that
the $\tilde {t}_{\bm \nu } $-copula distribution has the following
explicit integral expression
\begin{equation}
C_{\rm {\bm \nu }}^{\rm {\bf \Sigma }} ({\rm {\bf u}}) =
\int\limits_0^1 {\Phi _{\bm \Sigma} (z_1 (u_1 ,s),\dots,z_n (u_n
,s))ds}\label{cdfEqn}
\end{equation}
\noindent and its density is
\begin{equation}
c_{\rm {\bm \nu }}^{\rm {\bm \Sigma }} ({\rm {\bf u}}) =
\frac{\partial ^nC_{\rm {\bm \nu }}^{\rm {\bf \Sigma }} ({\rm {\bf
u}})}{\partial u_1 \dots\partial u_n } = \int\limits_0^1 {\varphi
_{\rm {\bf \Sigma }} \left( {z_1 (u_1 ,s),\dots,z_n (u_n ,s)}
\right)
 } \prod\limits_{i = 1}^n {[w_i (s)]^{ - 1}} ds / \prod\limits_{i =
1}^n {f_{\nu _i } (x_i )} . \label{densityEqn}
\end{equation}
\noindent Here:
\begin{itemize}

\item $ z_i (u_i ,s) = t_{\nu _i }^{ - 1} (u_i ) / w_i (s),\;i =
1,2,\ldots,n$;

\item $w_i (s) = G_{\nu _i }^{ - 1} (s)$;

\item $\varphi _{\rm {\bf \Sigma }} (z_1 ,\dots,z_n ) = \exp ( -
\textstyle{1 \over 2}{\rm {\bf {z}'\Sigma }}^{ - 1}{\rm {\bf z}}) /
[(2\pi )^{n / 2}(\textstyle{\det} {\rm {\bf \Sigma }})^{1 / 2}]$ is
the multivariate normal density;

\item $x_i = t_{\nu _i }^{ - 1} (u_i ),\;i = 1,2,\ldots,n$;

\item $f_\nu (x) = \left( {1 + x^2 / \nu } \right)^{ - (\nu+1)/2 }\Gamma ((\nu + 1)/2) / [\Gamma (\nu/2 )\sqrt {\nu \pi } ]$ is the
univariate $t$-density, where $\Gamma(\cdot)$ is a gamma function.

\end{itemize}

The multivariate density (\ref{densityEqn})  involves a
one-dimensional integration which makes the density calculation
computationally more demanding than in the case of the standard
$t$-copula, but still practical using available fast and accurate
algorithms for the one-dimensional integration. If all the dof
parameters are equal, i.e. $\nu _1 = \dots = \nu _n = \nu $, then it
is easy to show that the copula defined by (\ref{cdfEqn}) becomes
the standard $t$-copula; see Luo and Shevchenko (2010) for a proof.

\subsection{FX data and GARCH filtering}
\label{subsec:mylabel1}

 As a case study we consider modeling dependence between six FXs
 using $t$-copulas introduced in previous section.
The daily foreign exchange rate data for the six FX majors in the
period January 2004 to April 2008 (a total of 1092 trading days)
were downloaded from the \textit{Federal Reserve Statistical Release
}(\underline {http://www.federalreserve.gov/releases}). These daily
data have been certified by the Federal Reserve Bank of New York as
the noon buying rates in New York City. For our purpose, we study
the six major currencies (AUD, CAD, CHF, EUR, GBP and JPY). Rates
were converted to USD per currency unit in the present study, if not
already in this convention. This unified convention allows a
portfolio of currencies to be conveniently valued in terms of a
single currency, the USD.

Following common practice (see \nocite{McFrEm05} McNeil
 \textit{et al} 2005), we use the GARCH(1,1) model
to standardize the log-returns of the exchange rates marginally. The
GARCH(1,1) model calculates the current squared volatility $\sigma
_t^2 $ as
\begin{equation}
\sigma _t^2 = \omega + \alpha {\kern 1pt} (x_{t - 1}-\mu )^2 + \beta
\sigma _{t - 1}^2 ,\quad \omega \ge 0,\,\;\alpha ,\beta \ge
0,\;\alpha + \beta < 1,
\end{equation}
\noindent where $x_{t - 1} $ denotes the log-return of an exchange
rate on date $t - 1$. GARCH parameters $\omega $, $\alpha $ and
$\beta $ are estimated using the maximum likelihood method.
Log-return was modeled as
\begin{equation}
x_t = \mu + \sigma _t \varepsilon ^{(t)},
\end{equation}
\noindent where $\mu $ is the average historical return or drift for
the asset and $\varepsilon ^{(t)}$ is a sequence of iid random
variables referred to as the residuals. The GARCH filtered residuals
of the FX rates were then used to fit the $t$-copula models. Before
the  fitting the residuals were transformed to the (0,1) domain
marginally using empirical distributions of the residuals.

\subsection{Configuration of grouped t-copula}
\label{subsec:configuration}

With six dimensions, the grouped $t$-copula can have a total of 201
possible combinations (not counting the standard $t$-copula and the
 $\tilde {t}_{\bm \nu } $-copula). In this study we
concentrate on the class of configurations with two groups only,
which is the next level of complexity compared with the standard
$t$-copula. This reduces the number of possible grouped $t$-copula
models to 31. These 31 grouped  $t$-copula models are:

\begin{itemize}
\item 10 models from the complete subset of (3,3) configurations (with two
groups and three members in each group).
\item 15 models from the complete subset of (2,4) configurations (with two
members in the first group and four members in the second group).
\item  6 models from the complete subset of (1,5) configurations
(with one member in the first group and five members in the second
group).
\end{itemize}

Note, a (1,5) combination is the same as a (5,1) combination, and a
(2,4) combination is the same as a (4,2) combination. So, altogether
we have 33 competing models to choose from -- the standard
$t$-copula, the 31 two-grouped $t$-copula and the generalized
$t$-copula ($\tilde {t}_{\bm \nu } $-copula).

Table 1 lists all 33 models for modeling the six FX majors, their
grouping configurations and parameter notations. In column 2 of
Table 1, each pair of parentheses define a sub-group configuration.
The generalized grouped $t$-copula has six sub-groups with a single
member in each sub-group, while the standard $t$-copula has one
group containing all  six members. Note that for the grouped
$t$-copula, exchanging the two sub-groups makes no difference --
these two configurations have exactly the same combinations of
members, so no new models will emerge from this exchange.

\subsection{Maximum likelihood calibration}

Consider a random vector of data ${\rm {\bf Y}} = (Y_1 ,\dots,Y_n
{)}'$. To estimate a parametric copula using observations ${\rm {\bf
y}}^{(j)}$, $j = 1,\dots,K$, where $K$ is the number of
observations, the first step is to project the data to the $[0,1]^n$
domain to obtain ${\rm {\bf u}}^{(j)}$, using estimated marginal
distributions. In our study the margins are modeled  using empirical
distributions but it  can also be modeled using parametric
distributions or a combination of these methods, e.g. empirical
distribution for the body and a generalized Pareto distribution for
the tail of a marginal distribution (McNeil et al 2005, page 233).
Given pseudo sample ${\rm {\bf u}}^{(j)}$ constructed using the
original data, the copula parameters can be estimated using, for
example, the maximum likehood method or MCMC.

Accurate maximum likelihood estimates (MLEs) of the copula
parameters should be obtained by fitting all unknown parameters
jointly. In practice, to simplify the calibration procedure,
correlation matrix coefficients for $t$-copulas are often calculated
pair-wise using Kendall's \textit{tau} rank correlation coefficients
$\tau (Y_i ,Y_j )$ via the formula (McNeil \textit{et al} 2005)
\begin{equation}
\Sigma _{ij} = \sin \left( {\textstyle{1 \over 2}\pi \tau (Y_i ,Y_j
)} \right).\label{tauEqn}
\end{equation}
Then in a second stage the dof parameters $\nu_1,\dots,\nu_n$ are
estimated. Strictly speaking (\ref{tauEqn}) is valid for bivariate
case only, however in practice it works well for multivariate case
too. It was noted in Daul \textit{et al} (2003) that formula (12) is
still highly accurate even when it is applied to find the
correlation coefficients between risks from the different groups.
McNeil {\it et al} (2005) observed that the estimated parameters
using Kendall's \textit{tau} are identical to those obtained by
joint estimation to two significant digits, confirming good accuracy
of the Kendall's \textit{tau} simplification. It was also observed
in Luo and Shevchenko (2010) that the difference in estimated
parameters between the Kendall's \textit{tau} approximation and the
joint estimation was mostly in the third significant digit and was
smaller than the standard errors for the MLEs. In addition, a study
of small sample properties in Luo and Shevchenko (2010) showed that
the bias introduced by the Kendall's \textit{tau} approximation is
very small even for a small sample size of 50. In the present work
the data sample size is over 1000. The small bias of the Kendall's
\textit{tau} approximation is certainly insignificant when compared
with the often large difference existing between dof parameters of
different $t$-copula models. In other words, using (\ref{tauEqn})
for the correlation coefficients should cause little material
difference in the present model choice study where the difference is
expected to come from different group configurations.

Because the Kendall's \textit{tau} approximation is applied
pair-wise, we have identical correlation matrix for all the copula
models to be considered. This simplification is computationally very
significant for the grouped $t$-copula for which the calibration
using density (\ref{densityEqn})  is computationally demanding. By
using the Kendall's \textit{tau} approximation, the number of
unknown parameters reduces from $M = n(n + 1) / 2$ to $M = n$ for
the generalized grouped $t$-copula. With six-dimensions considered
in this study, this amounts to a reduction from 21 parameters to
only 6. For the grouped $t$-copula with two groups, this reduction
is from 17 to 2, an even more dramatic reduction. A substantial
saving of computing time is achieved in both cases.

\vspace{0.2cm}

 \noindent {\bf Remark:} An  accurate calibration of grouped $t$-copula requires joint estimation
 of dof parameters. Sometimes in practice an approximate approach is taken where
 a grouped $t$-copula is  calibrated marginally, i.e. each sub-group is
 calibrated separately using a standard $t$-copula. This approximation is not always
 justified; also it can not be applied to a generalized $t$-copula.
  For a proper and fair comparison between the grouped $t$-copula and the generalized grouped $t$-copula,
  in this study we perform joint calibration for both copulas.
   When the grouped $t$-copula is calibrated
jointly, its density is given by the integral formula
(\ref{densityEqn}), the same as the generalized $t$-copula, so a
proper joint calibration of the grouped $t$-copula is also
computationally demanding when compared with the calibration of a
standard $t$-copula.

 Let ${\bm \nu}$ be the vector of
$n$ dof parameters $\nu_1,\dots,\nu_n$ (the grouped $t$-copula is
treated as a special case of $\tilde {t}_{\bm \nu } $-copula).
Denote the density of the $\tilde {t}_{\bm \nu } $-copula  evaluated
at ${\rm {\bf u}}^{(j)}$ as $c_{\bm \nu } ({\rm {\bf u}}^{(j)})$,
which can be obtained using (\ref{densityEqn}). Then the  MLEs for
${\bm \nu }$ are calculated by maximizing the log-likelihood
function
\begin{eqnarray}
\ell_{\rm {\bf U}} ({\bm {\nu }}) &=&  \ln \prod\limits_{j = 1}^K
{c^\Sigma_{\bm {\nu }}
({\rm {\bf u}}^{(j)})} \nonumber\\
 &=&  \sum\limits_{j = 1}^K {\ln \left( {\int\limits_0^1 {\varphi_{\rm
{\bf \Sigma }} \left( {z_1^{(j)}(s),\ldots,z_n^{(j)}(s)} \right)}
\prod\limits_{i = 1}^n {[w_i (s)]^{ - 1}} ds} \right)}
\nonumber \\
&& + \sum\limits_{j = 1}^K {\sum\limits_{i = 1}^n {\left(
{\textstyle{1 \over 2}(\nu _i + 1)} \right)\ln [1 + (x_i^{(j)} )^2 /
\nu _i ]} }\nonumber\\
&& + K\sum\limits_{i = 1}^n {\left( {\textstyle{1 \over 2}\ln (\nu
_i \pi ) + \ln [\Gamma (\textstyle{1 \over 2}\nu _i ) / \Gamma
(\textstyle{1 \over 2}(\nu _i + 1))]} \right)},\label{logLeqn}
\end{eqnarray}
\noindent where $x_i^{(j)} = t_{\nu _i }^{ - 1} (u_i^{(j)} )$,
$z_i^{(j)}(s)=x_i^{(j)} / w_i (s)$, $i = 1,\ldots,n$, $j =
1,\ldots,K$. In this work we use the double precision \textit{IMSL}
function DQDAGS, a globally adaptive integration scheme documented
in \nocite{PiDoKaUbKa83} Piessens \textit{et al} (1983) for the
integration in (\ref{densityEqn}). For the maximization of
(\ref{logLeqn}) the double precision \textit{IMSL} function DBCPOL
is used, which employs a direct search Simplex algorithm that does
not require calculation of gradients.

\bigskip

\section{Bayesian inference and MCMC}
\label{sec:bayesian}

In this section we describe Bayesian approach and MCMC procedure to
estimate $t$-copulas, and model selection criteria used to choose
the $t$-copula model. Under the Bayesian approach, the model
parameters ${\bm \theta }$ (in our case ${\bm \theta }$ is just the
dof parameter ${\bm \nu }$) are treated as random variables. Given a
\textit{prior} distribution $\pi ({\bm \theta })$ and a conditional
density of the data given ${\bm \theta }$ (i.e. likelihood) $\pi
({\rm {\bf y}}\vert {\bm \theta })$, the joint density of data ${\rm
{\bf Y}}$ and the model parameters ${\bm \theta }$ is $\pi ({\rm
{\bf y}},{\bm \theta }) = \pi ({\rm {\bf y}}\vert {\bm \theta })\pi
({\bm \theta })$. Having observed data ${\rm {\bf Y}}$, the
distribution of ${\bm \theta }$ conditional on ${\rm {\bf Y}}$, the
\textit{posterior} distribution, is determined by  Bayes' theorem
\begin{equation}
\pi ({\bm \theta }\vert {\rm {\bf y}}) = \frac{\pi ({\rm {\bf
y}}\vert {\bm \theta })\pi ({\bm \theta })}{\int {\pi ({\rm {\bf
y}}\vert {\bm \theta })\pi ({\bm \theta })d{\bm \theta }} }\propto
\pi ({\rm {\bf y}}\vert {\bm \theta })\pi ({\bm \theta
}).\label{bayesEqn}
\end{equation}
The posterior can then be used for predictive inference.  There is a
large number of useful texts on Bayesian inference; for a good
introduction, see \nocite{Berger85} Berger (1985) and \nocite{Rob01}
Robert (2001).

\bigskip

\subsection{MCMC under Bayesian framework}

The explicit evaluation of the normalization constant in
(\ref{bayesEqn}) is often difficult especially in high dimensions.
The complexity in our case is evident from the log-likelihood
expression (\ref{logLeqn}). The MCMC method provides a highly
efficient alternative to traditional techniques by sampling from the
posterior indirectly and performing the integration implicitly.

MCMC is especially suited to a Bayesian inference framework. It
facilitates the quantification of parameter uncertainty and model
risks. It also allows a unified estimation procedure that estimates
parameters and latent variables. In the last case a special
algorithm called data augmentation can be employed, see
\nocite{TannerWong87} Tanner and Wong (1987).   The Bayesian
estimates of particular interest from MCMC are the maximum \textit{a
posterior} (MAP) estimate and the minimum mean square error (MMSE)
estimate, defined as follows
\begin{equation}
\mbox{MAP}:\quad {\hat {\bm \theta }}^{MAP} = \arg \mathop {\max
}\limits_{\bm \theta } [\pi ({\bm \theta }\vert {\rm {\bf y}})],
\end{equation}
\begin{equation}
\mbox{MMSE}:\quad {\hat {\bm \theta }}^{MMSE} = E[{\bm \theta }\vert
{\rm {\bf y}}].
\end{equation}
The MAP and MMSE estimates are the posterior mode and mean
respectively. If the prior $\pi ({\bm \theta })$ is constant and the
parameter range includes the MLE, then the MAP of the posterior is
the same as MLE.

\subsection{Metropolis-Hastings algorithm}
\label{subsec:metropolis}

In our case study we use the Metropolis-Hastings algorithm first
described by Hastings (1970) as a generalization of the Metropolis
algorithm (\nocite{MetRT53}Metropolis {\it et al} 1953). Denote the
state vector at step $t$ as ${\bm \theta }^{(t)}$ and we wish to
update it to a new state ${\bm \theta }^{(t + 1)}$.  We generate a
candidate ${\bm \theta }^
* $ from density $q( {\bm \theta } \vert {\bm \theta }^{(t)})$, and accept this point as the new state of the chain with
probability given by
\begin{equation}
\alpha ({\bm \theta }^{(t)},{\bm \theta }^ * ) = \min \left\{
{1,\;\frac{\pi({\bm \theta }^ * )q({\bm \theta }^{(t)}
 \vert {\bm \theta }^{*})}{\pi({\bm \theta }^{(t)})q({\bm \theta }^{*} \vert {\bm \theta }^{(t)} )}} \right\}.
\label{acceptEqn}
\end{equation}
If the proposal is accepted,  the new state ${\bm \theta }^{(t + 1)}
= {\bm \theta }^ * $, otherwise ${\bm \theta }^{(t + 1)} = {\bm
\theta }^{(t)}$.  The single component Metropolis-Hastings is often
more efficient in practice. Here  the state variable ${\bm \theta
}^{(t)}$
 is partitioned into components ${\bm \theta }^{(t)} = (\theta _1^{(t)}
,\theta _2^{(t)} ,\dots,\theta _n^{(t)} )$ which are updated one by
one or block by block. This was the framework for MCMC originally
proposed by Metropolis \textit{et al}. (1953), and is adapted in
this study.

The likelihood is computed as $\pi ({\rm {\bf y}}\vert {\bm \theta
}) = \exp (\ell_{\bm y} ({\bm \theta }))$, where $\ell_{\bm y} ({\bm
\theta })$ is the log-likelihood given by (\ref{logLeqn}). In
computer implementation, we take advantage of the fact that only one
component is updated at each sub-step in the single component
Metropolis-Hastings algorithm by saving and re-using any values not
affected by the current updating. For example, each evaluation of
(\ref{logLeqn}) calls for the inverse of the $t$-distribution for
all the data points and all the dof values. Saving and re-using
these inverse values reduce the calculation by a factor of six for
the six-dimensional MCMC computation.

\subsection{Bayesian model selection using MCMC}

Powerful MCMC methods such as the Gibbs sampler
(\nocite{GelS90}Gelfand and Smith 1990) and the Metropolis-Hastings
(MH) algorithm (\nocite{Has70}Hastings 1970) enable direct
estimation of the posterior and predictive quantities of interest,
but do not lend themselves readily to estimation of the model
probabilities. While one of the most common classical techniques is
the Bayesian Information Criterion (BIC) (\nocite{Sch78}Schwarz
1978), many new approaches have been suggested in the literature.

The most widely used methods include the harmonic mean estimator of
\nocite{NewR94} Newton and Raftery (1994), importance sampling
(\nocite{Fru95}Fruhwirth-Schnatter 1995), the reciprocal importance
sampling estimator (\nocite{GelD94}Gelfand and Dey 1994), and bridge
sampling (\nocite{MenW96}Meng and Wong 1996, \nocite{Fru04}
Fruhwirth-Schnatter 2004). A comprehensive review of some of these
methods applied to Bayesian model selection can be found in Kass and
Raftery \nocite{KasR95} (1995).

 Consider model $M$ with parameter vector ${\bm
\theta }$. The model likelihood with data ${\rm {\bf y}}$ can be
found by integrating out the parameter ${\bm \theta }$
\begin{equation}
\pi({\rm {\bf y}}\vert M) = \int {\pi({\rm {\bf y}}\vert {\bm \theta
},M)} \pi({\rm {\bm \theta }}\vert M)d{\bm \theta
},\label{bayesIntEqn}
\end{equation}
\noindent where $\pi({\bm \theta }\vert M)$ is the prior density of
${\bm \theta }$ in model $M$. Given a set of $H$ competing models
${\rm {\bf M}} = (M_1 ,M_2 ,\dots,M_H )$, the Bayesian alternative
to traditional hypothesis testing is to evaluate and compare the
posterior probability ratio between the models. For model $M_l$
($1\leq l \leq H$), assuming we have some prior knowledge about the
model probability $\pi (M_l )$, we can compute the posterior
probabilities for all models using the model likelihoods
\begin{equation}
\pi(M_l \vert {\rm {\bf y}}) = \frac{\pi({\rm {\bf y}}\vert M_l
)\;\pi (M_l )}{\sum\nolimits_{h = 1}^H {\pi({\rm {\bf y}}\vert M_h
)\;\pi (M_h )} }.
\end{equation}
Consider two competing models $M_1 $ and $M_2 $, parameterized by
${\bm \theta }_{[1]}$ and ${\bm \theta }_{[2]}$ respectively. The
choice between the two models can be based on the posterior model
probability ratio, given by
\begin{equation}
\frac{\pi(M_1 \vert {\rm {\bf y}})}{\pi(M_2 \vert {\rm {\bf y}})} =
\frac{\pi({\rm {\bf y}}\vert M_1 )\;\pi (M_1 )}{\pi({\rm {\bf
y}}\vert M_2 )\;\pi (M_2 )} = \frac{\pi (M_1 )}{\pi (M_2 )}B_{12},
\label{bayesF}
\end{equation}
\noindent where $B_{12} = \pi({\rm {\bf y}}\vert M_1 ) / \pi({\rm
{\bf y}}\vert M_2 )$ is the Bayes factor, the ratio of posterior
odds of model $M_1 $ to that of model $M_2 $. As shown by
\nocite{LavS99} Lavine and Scherrish (1999), an accurate
interpretation of the Bayes factor is that the ratio $B_{12} $
captures the change of the odds in favour of model $M_1 $ as we move
from prior to posterior. \nocite{Jef61} Jeffreys (1961) recommended
a scale of evidence for interpreting Bayes factors, which was later
modified by \nocite{Was97} Wasserman (1997). A Bayes factor $B_{12}
> 10$ is considered strong evidence in favour of $M_1 $. For a detailed review of Bayes factors,
see Kass and Raftery (1995).

Typically, the integral (\ref{bayesIntEqn}) required by the Bayes
factor is not analytically tractable and sampling based methods must
be used to obtain estimates of the model likelihoods. In the current
study we choose three  methods for model selection:

\begin{itemize}
\item direct estimation of the Bayes factor in (\ref{bayesF}) using Reciprocal Importance
Sampling Estimation presented in Section \ref{riseSec};
\item deviance information criterion (see Section \ref{dicSec});
\item direct computation of the posterior model probabilities using formula presented in Section \ref{pmpSec}.
\end{itemize}

\subsubsection{Reciprocal Importance Sampling
Estimator}\label{riseSec}

Given samples ${\bm \theta }^{(t)},\;t = 1,\ldots,N$ from the
posterior distribution obtained through MCMC, Gelfand and Dey (1994)
proposed the \textit{reciprocal importance sampling estimator
}(RISE) to approximate the model likelihood as
\begin{equation}
\pi ({\rm {\bf y}}|M) \approx \left[ {\frac{1}{N}\sum\limits_{t =
1}^N {\frac{h({\bm \theta }^{(t)} \vert M)}{\pi({\rm {\bf y}}\vert
{\bm \theta }^{(t)},M)\;\pi ({\bm \theta }^{(t)}\vert M )}} }
\right]^{ - 1},\label{riseEqn}
\end{equation}
\noindent where $h$ plays the role of an importance sampling density
roughly matching the posterior. Gelfand and Dey (1994) suggested a
multivariate normal or $t$-distribution density with mean and
covariance fitted to the posterior sample.

The RISE estimator can be regarded as a generalization of the
\textit{harmonic mean estimator }suggested by Newton and Raftery
(1994). If $h = 1$ then (\ref{riseEqn}) becomes the harmonic mean
estimator. Other estimators include the \textit{bridge sampling}
proposed by Meng and Wong (1996), and the \textit{Chib's candidate's
estimator} (Chib 1995). \nocite{Chi95} In a recent comparison study
by Miazhynskaia and Dorffner (2006), \nocite{MiaD06} these
estimators were employed as competing methods for Bayesian model
selection on GARCH-type models, along with the reversible jump MCMC.
It was demonstrated that the RISE estimator (either with normal or
$t$ importance sampling density), the bridge sampling method and the
Chib's algorithm gave statistically equal performance in model
selection, and their performance more or less matches the much more
involved reversible jump MCMC.

\subsubsection{Deviance Information Criterion }\label{dicSec}
\noindent The deviance information criterion (DIC) is a
generalization of the \textit{Bayesian information criterion}
(Schwarz 1978, \nocite{spi2002} Spiegelhalter et al 2002). For a
given model $M$ (for simplicity we drop notation $M$ in the formula
below) the deviance is defined as
\begin{equation}
D({\bm \theta }) = - 2\log (\pi({\rm {\bf y}}\vert {\bm \theta })) +
C, \label{dicEqnC}
\end{equation}
\noindent where the constant $C$ is common to all nested models.
Then DIC is calculated as
\begin{equation}
DIC = 2E_{\rm {\bf \theta }} [D({\bm \theta })] - D(E_{\bm \theta }
[{\bm \theta }]) = E_{\bm \theta } [D({\bm \theta })] + (E_{\bm
\theta } [D({\bm \theta })] - D(E_{\bm \theta } [{\bm \theta
}])),\label{dicEqn}
\end{equation}
\noindent where $E_{\bm \theta } [\cdot]$ is the expectation with
respect to ${\bm \theta }$. The expectation $E_{\bm \theta } [D({\bm
\theta })]$ is a measure of how well the model fits the data; the
smaller its value, the better the fit. The difference $E_{\rm {\bf
\theta }} [D({\bm \theta })] - D(E_{\bm \theta } [{\bm \theta }])$
can be regarded as the effective number of parameters, the larger
this term, the easier it is for the model to fit the data. So the
DIC criterion favours the model with a better fit but at the same
time penalizes the model with more parameters. Under this setting
the model with the smallest DIC value is the preferred model.

\subsubsection{Posterior model probabilities }\label{pmpSec}

A popular approach for model choice is based on Reversible Jump MCMC
(Green 1995). \nocite{Gre95} Here we adopt an alternative proposed
recently by Peters {\it et al} (2009) based on the work of
\nocite{Con06} Congdon (2006). In this procedure the posterior model
probabilities $\pi(M_l \vert {\rm {\bf y}})$ are estimated using the
Markov chain in each model as
\begin{equation}
\pi(M_l \vert {\rm {\bf y}}) = \sum\limits_{t = 1}^N {\;\frac{L_{\rm
{\bf y}}(M_l ,{\bm \theta }_{[l]}^{(t)} )\;}{\sum\nolimits_{h = 1}^H
{L_{\rm {\bf y}}(M_h ,{\bm \theta }_{[h]}^{(t)} )\;}
}},\label{postProbEqn}
\end{equation}
\noindent where ${\bm \theta }_{[l]}^{(t)} $ is the MCMC posterior
sample at Markov chain step $t$ for model $M_l$, $L_{\rm {\bf
y}}(M_l ,{\bm \theta }_{[l]}^{(t)} )$ is the likelihood of ${\rm
{\bf y}}$ for a given model $M_l$ with parameter vector ${\bm \theta
}_{[l]}^{(t)} $, and $N$ is the total number of MCMC steps after
burn-in period. In (\ref{postProbEqn}), it is assumed that priors
$\pi(\theta_{[l]} \vert M_l)$ and $\pi(M_l)$ are constant.

\section{MCMC simulation results and analysis}
\label{sec:mylabel1}

\noindent \textbf{\textit{Prior distributions}}. In all  MCMC
simulation runs, we assume a uniform prior for every model
parameter. The only subjective judgement we bring to the prior is
the support of the dof parameter. Denote the $k^{th}$ dof parameter
of the $h^{th}$ $t$-copula model as $\nu _k^{(h)} $ (see Table 1).
We impose a common lower and upper bounds for all dof components,
specifically $1 = \nu _{\min } < \nu _k^{(h)} < \nu _{\max } = 100$
. In our case study the support ($1,\;100$) for dof parameter of the
$t$-distribution should be sufficiently large to allow the posterior
to be implied mainly by the observed data. To make sure the range is
sufficiently large, we also tested a wider range of $(1,\;200)$ and
found no material difference in the results.

\vspace{0.3cm}
 \noindent \textbf{\textit{MCMC procedure}}. The
starting value for the Markov chain for each component is set to a
uniform random number drawn independently from the support $(\nu
_{\min } ,\;\nu _{\max } )$. In the single component
Metropolis-Hastings algorithm, we adopt a truncated Gaussian
distribution as the symmetric random walk proposal density for
$q(\cdot|\cdot)$ in (\ref{acceptEqn}). For each component, the mean
of the Gaussian density was set to the current state and the
variance was pre-tuned so that the acceptance rate is close to the
optimal level. For $d$-dimensional target distributions with iid
components, the asymptotic optimal acceptance rate has been reported
to be 0.234; see \nocite{GelGR97} Gelman et al (1997) and
\nocite{RoRo01} Roberts and Rosenthal (2001). In pre-tuning the
variances for all the components we set 0.234 as the target
acceptance rate. In addition, the Gaussian density was truncated
below $\nu _{\min } $ and above $\nu _{\max } $ to ensure each
proposal was drawn within the support for the parameters.
Specifically, for the $k^{th}$ component at chain step $t$, the
proposal density is
\begin{equation}
q_k (\theta ^
* \vert \theta_k^{(t)})
 = \frac{f_N (\theta ^
* ;  \theta _k^{(t)} ,\sigma _k )}{F_N (\nu _{\max
} ; \theta _k^{(t)} ,\sigma _k ) - F_N (\nu _{\min } ;  \theta
_k^{(t)} ,\sigma _k )},
\end{equation}
\noindent where $f_N (\cdot; \mu ,\sigma )$ and $F_N (\cdot; \mu
,\sigma )$ are the Gaussian density and distribution  functions
respectively, with  mean $\mu $ and standard deviation $\sigma $.

An independent Markov chain was run for each of the 33 models listed
in Table 1. Each run consists of three stages:
\begin{itemize}
\item Tuning - tune and adjust the proposal standard
deviation to achieve optimal acceptance rate for each component.
\item ``Burn-in'' -  samples from this period are discarded.
\item Posterior sampling - here the Markov chian is considered to
have converged to the stationary target distribution and samples are
used for model estimates.
\end{itemize}

Unless stated otherwise, we use a ``burn-in'' period of length
 $N_b = 20,000$. We then let the
chain run for an additional $N = 100,000$ iterations to generate the
posterior samples. Each step contained a complete update of all
components.

\vspace{0.3cm}
 \noindent \textbf{\textit{MCMC convergence}}. Figure
1 shows the first 30,000 samples, taken \textit{after} the burn-in
period, of the dof component $\nu^{(0)}_1$  for model $M_0 $ (i.e.
the case of the generalized  $t$-copula). Since $M_0 $ has the
highest parameter dimensions among all the candidate models, in
general it requires the longest length of chains to converge to a
stationary distribution. This figure shows that after the burn-in
period the samples are mixing well over the support of the posterior
distribution.

In addition to inspecting the sample paths, we also monitor the
autocorrelation of the samples. Figure 2 shows the autocorrelations
over multiple lags computed from the posterior samples for component
$\nu^{(0)}_1$  of model $M_0 $. A useful value to compute from these
autocorrelations for each component is the autocorrelation time
 defined as
\begin{equation}
\tau _k = 1 + 2\sum\limits_{g = 1}^\infty {\rho _k
(g)},\label{sumRhoEqn}
\end{equation}
\noindent where $\rho _k (g)$ is the autocorrelation at lag $g$ for
component $\theta _k $. This autocorrelation  is sometimes used to
compute an ``effective sample size'' by dividing the number of
samples by $\tau_k $. The standard errors for the parameters can
then be based on the effective sample size to compensate for the
autocorrelation (see \nocite{Rip87} Ripley 1987, \nocite{Nea93} Neal
1993). In practice it is necessary to cut off the sum in
(\ref{sumRhoEqn}) at  $g = g_k^{\max } $ where the autocorrelations
seem to have fallen to near zero, because including  higher lags
adds too much noise (for some interesting discussion on this issue,
see  \nocite{KasCG98} Kass \textit{et al}. 1998). As shown in Figure
2, in those well mixed MCMC samples the autocorrelation falls to
near zero quickly and stays near zero at larger lags. For this study
we have chosen  a $g_k^{\max } $ for each component such that the
autocorrelation at lag $g_k^{\max } $ has reduced to less than 0.01.
That is, the autocorrelation  time $\tau _k $ is estimated by
\begin{equation}
\hat {\tau }_k \approx 1 + 2\sum\limits_{g = 1}^{g_k^{\max } } {\rho
_k (g)} ,\quad g_k^{\max } = \min \{g:\rho _k (g) < 0.01\}.
\end{equation}
The $\hat {\tau }_k $ values estimated from MCMC output for model
$M_0 $ are shown in Table 2, along with the cut-off lag number
$g_k^{\max } $. MCMC convergence characteristics for other
components and for other models are very similar to those shown here
for model $M_0 $.

\subsection{Bayesian estimates of parameters}
\label{subsec:bayesian} \noindent This section presents  results for
posterior mean (MMSE), mode (MAP) and numerical error due to finite
number of MCMC iterations.

\subsubsection{Posterior mean and its numerical error}

Table 3 shows values for the estimated mean from MCMC posterior
samples for all 33 models. The standard errors (numerical error due
to finite number of MCMC iterations) are shown in parentheses and
the log-likelihoods corresponding to the estimated means are in the
last column. Since the samples from MCMC are typically serially
correlated, the usual formula for estimating the standard error of a
sample mean (i.e. standard deviation divided by $\sqrt{N}$) will
introduce significant under-estimation. Here, we use \textit{batch
sampling} for the standard error estimate of the MCMC posterior
mean; see \nocite{Gilks96} Gilks \textit{et al}. (1996).

Consider a MCMC posterior sample $y_1,y_2,...,y_N$ with length $N =
Q\times L$, where $L$ is sufficiently large, so that the batch means
\begin{equation}
\bar {y}_q = \frac{1}{L}\quad \sum\limits_{t = (q - 1)L +
1}^{q\times L} {\quad y^{(t)}\quad ,\quad q = 1,\dots,Q}
\end{equation}
\noindent are considered approximately independent. Then $\bar
{y}=(\bar{y}_1+\dots+\bar{y}_Q)/Q$ and the standard error of the
posterior sample mean $\bar {y}$ can be approximated by
\begin{equation}
\sqrt{\mathrm{Var}(\bar{y})} \approx \frac{1}{\sqrt{Q}}\sqrt
{\frac{1}{Q - 1}\quad \sum\limits_{q = 1}^Q {(\bar {y}_q - \bar
{y})^2}},
\end{equation}
\noindent  Note that $Q$ is the number of quasi-independent batches
 and $L = N / Q$ is the size of each  batch.

\subsubsection{Posterior mode and likelihood ratio tests}
\label{subsubsec:posterior}

Values of the posterior mode taken from the MCMC samples for all the
33 models are shown in Table 4, along with the corresponding
log-likelihood values. Using results of the maximum likelihood
corresponding the posterior mode in Table 4, a classical likelihood
ratio test  can be performed to compare  model likelihoods.

Consider the null hypothesis that the observed FX daily return data
are from distribution described by the grouped $t$-copula model $M_1
$, and the alternative hypothesis that the data are distributed
according to the generalized  $t$-copula model $M_0 $. The
likelihood ratio for the two models is simply $\Lambda = L_1 / L_0
$, where $L_1 $ and $L_0 $ are the maximum likelihood values (i.e.
the likelihood value at the mode) for $M_1 $ and $M_0 $
respectively. The test statistic $ - 2\log (\Lambda )$ will be
asymptotically $\chi ^2$ distributed with degrees of freedom equal
to the difference in the number of dof parameters in $M_0 $ and $M_1
$, which is 4 in this case.

We can perform likelihood ratio tests on all  other grouped
$t$-copula models ($\mbox{M}_h ,\;h = 2,\dots,31$) against the same
alternative hypothesis of model $M_0 $, the generalized $t$-copula.
For the standard $t$-copula the difference in the number of
parameters is 5. The test statistic and the associated p-value
($\chi ^2$ significance) are given in Table 4. Clearly, according to
the p-value, all the null hypotheses should be rejected and the
alternative hypothesis, the generalized $t$-copula model $M_0 $, is
statistically justified.

Excluding model $M_0 $, among the other 32 $t$-copula models ($M_h
,\;h = 1,\dots,32$), the one achieving the highest likelihood is
$M_{27}$,
 which is one of the six (1,5) two-group configurations. The p-value of this
best grouped $t$-copula model against $M_0 $ is 0.0045 which is
still very small, suggesting a rather strong rejection of the
grouped $t$-copula (including the standard $t$-copula) in favour of
the $\tilde {t}_{\rm {\bf \nu }} $-copula model $M_0 $. Achieving
the highest likelihood from the fifteen (2,4) configurations is
model $M_4 $. It is interesting to notice that both  $M_{27} $ and
$M_4 $ have
 three European currencies (CHF, EUR, GBP) in one group (see Table 1),
 perhaps reflecting a natural geopolitical and economic grouping.

\subsection{Bayesian model choice}
\label{subsec:mylabel2}

While the likelihood ratio test relies on a single point estimate,
the Bayesian model choice makes decisions based on the entire
posterior distribution. As discussed in Section 3.3, three Bayesian
inference criteria were used to choose among the 33 $t$-copula
models: RISE given by (\ref{riseEqn}); DIC given by (\ref{dicEqn});
and the posterior model probabilities (\ref{postProbEqn}).

The RISE calculation involves fitting the MCMC posterior samples to
a multivariate normal or $t$-distribution and taking expectation of
the reciprocal likelihood. The DIC calculation requires taking
expectation of the likelihood and the parameters. Column 2 in Table
5 shows the RISE  factor $B_{0h} = R_0 / R_h ,\;h = 1,\dots,32$,
where $R_h $ is the RISE value for model $M_h $. That is, $B_{0h} $
($1 \le h \le 32)$ is a measure of strength in the argument that the
generalized  $t$-copula (model $M_0 $) is the Bayesian choice. The
very large Bayes factors ($B_{0h} > e^{11}
> 5.9\times 10^4)$ shown in Table 5 overwhelmingly support the
generalized  $t$-copula, confirming the likelihood ratio tests
discussed previously. Excluding model $M_0 $, these Bayes factors
also point to $M_{27} $  as the most favoured model among the
grouped $t$-copulas ($M_h $, $1 \le h \le 32$) confirming the
likelihood ratio tests. The larger the Bayes factor $B_{0h} $, the
stronger the case against model $M_h $ ($1 \le h \le 32)$.

The DIC values for 33 models are shown in column 3 of Table 5. Since
only relative DIC value matters, the common constant in
(\ref{dicEqnC}) was set in such a way that the DIC value for model
$M_0 $ is zero. As shown in Table 5, the DIC value for all the other
models is significantly positive, relative to that of $M_0 $. Thus
under the DIC criterion the model of choice is clearly $M_0 $, i.e.
the generalized $t$-copula. In addition, similar to the RISE based
Bayes factors and the likelihood tests, the DIC values also pick
$M_{27} $ as the most likely grouped $t$-copula model after $M_0 $,
the same as the RISE factor and the likelihood ratio test. The
larger the DIC value, the stronger the case  against the model. It
is interesting to observe that the magnitude of the DIC value is
close to that of the logarithm of the Bayes factor based on the
reciprocal importance sampling estimator, when both are evaluated
relative to the same model $M_0$.

As shown by column 4 in Table 5, the results for posterior model
probabilities (\ref{postProbEqn}) also agree with the RISE and DIC
results, i.e. model $M_0 $ has a very high probability of $88\%$,
and model $M_{27} $ has the second highest probability. If we
exclude model $M_0 $, then model $M_{27}$ has a high probability of
$68\%$. In summary, all three Bayesian choice criteria point to the
same model $M_0 $ as the best choice followed by model $M_{27} $,
and these choices are  in agreement with the classical likelihood
ratio tests, as shown in Table 4.

\section{Conditional Value-at-Risk}
\label{sec:conditional}

Consider a portfolio of six major currencies. Denote the exchange
rates (USD per currency unit) for these currencies at time $t$ by
$S_i^{(t)} ,\;i = 1,\dots,6$. Assume we hold $\lambda _i $ units for
the $i^{th}$ currency. The portfolio value at time $t$ is then
$V^{(t)} = \sum\nolimits_{i = 1}^6 {\lambda _i S_i^{(t)} } $. The
log-return for the $i^{th}$ currency at time $t + 1$ is given by
$x_i^{(t + 1)} = \ln S_i^{(t + 1)} - \ln S_i^{(t)} $. The portfolio
loss for one time step is then
\begin{eqnarray}
- \delta V^{(t + 1)} = V^{(t)} - V^{(t + 1)} &=& \sum\limits_{i =
1}^6 {\lambda _i S_i^{(t)} \left( {1 - \exp \left( {x_i^{(t + 1)} }
\right)} \right)}  \nonumber \\
&=&  V^{(t)}\sum\limits_{i = 1}^6 {w_i \left( {1 -\exp \left(
{x_i^{(t + 1)} } \right)} \right)},
\end{eqnarray}

\noindent where $w_i = \lambda _i S_i^{(t)} / V^{(t)}$ is the
proportion of the portfolio value in currency $i$ at time $t$, that
is, it is the dollar weight of the $i^{th}$ currency. Now we wish to
simulate the distribution of portfolio  return
$$
Z = - \delta V^{(t + 1)} / V^{(t)} = \sum\limits_{i = 1}^6 {w_i \left( {1 -
\exp \left( {x_i^{(t + 1)} } \right)} \right)} \approx \sum\limits_{i = 1}^6
{ - w_i x_i^{(t + 1)} } .
$$
In the present study we model the dependence of the log-returns
$x_i^{(t + 1)} $ by one of the $t$-copula models as described in the
previous sections. Recall that the dof parameters and their
posterior distributions are already obtained by Bayesian MCMC. To
focus on the impact of copula models, we use the standard normal
distribution for all the six marginals. We take the CVaR as our risk
measure. Assume that a random variable $Z$ has continuous density
$f( \cdot )$ and distribution $F( \cdot )$. Given a threshold
quantile level $\alpha $, the CVaR above $F^{ - 1}(\alpha )$ is
defined as
\begin{equation}
CVaR_\alpha [Z] = E[Z \vert Z\geq F^{ - 1}(\alpha )] = \frac{1}{1 -
\alpha }{\kern 1pt} \,\int\limits_{F^{ - 1}(\alpha )}^\infty
{xf(x)dx} {\kern 1pt},
\end{equation}
\noindent which is  the average of the losses exceeding $F^{ -
1}(\alpha )$. To demonstrate model impact on risk quantification, we
compare CVaR of the two most likely models,  $M_0 $ and $M_{27} $,
the best and the second best models of all  33 candidates. CVaR is
calculated numerically using $10^7$ Monte Carlo simulations with
$t$-copula model parameters given in Table (4).

Table 6 shows $CVaR_{0.99}^{(M_0 )} $ and $CVaR_{0.99}^{(M_{27} )} $
predicted by models $M_0 $ and $M_{27} $ for two portfolios (defined
by weights in Table 6). Note in both portfolios we have negative
weights (selling the currency) and the weights in each portfolio add
to 1.0. As shown in Table 6, model $M_{27} $ underestimates the 0.99
CVaR by 16{\%} for the first portfolio, and this underestimate
reverses to a slight overestimate for the second portfolio, assuming
the correct estimates are from model $M_0 $. The second portfolio is
only slightly different from the first -- swapping the position of
EUR and CHF (long/short position) in the first portfolio yields the
second portfolio. The two portfolios are deliberately chosen to
demonstrate that the model impact on risk quantification can be in
either direction -- it may be overestimation or it may be
underestimation, depending on the portfolio.
 Thus it is important to choose the most suitable model
statistically, such as by means of Bayesian model inference. Table 7
compares 0.99 CVaR prediction of model $M_0 $ with that of model
$M_4 $, the most likely model from the (3,3) configuration, for the
same two portfolios as those in Table 6. Here again the 0.99 CVaR
for the first portfolio is underestimated by the incorrect model,
and for the second portfolio it is overestimated.

\section{Conclusion}

This paper describes a Bayesian model choice methodology for
$t$-copula models. As an illustration, altogether 33 $t$-copula
models of six dimensions were considered: the generalized
$t$-copula; the standard $t$-copula; and  31 grouped $t$-copula
models from the complete subset of (3,3), (2,4) and (1,5)
configurations. MCMC simulations under a Bayesian inference
framework were performed to obtain the posterior distribution of dof
parameters for all  33 $t$-copula models. Using historical data of
foreign exchange rates as a case study, we found that Bayesian model
choice based on the RISE, the DIC
 and the \textit{posterior
model probabilities}  overwhelmingly favors the generalized
$t$-copula model $M_0 $. In addition, all three Bayesian choice
criteria  point to the same second most likely model $M_{27}$. These
Bayesian choices are also in agreement with  classical likelihood
ratio tests.

The impact of model choice on the CVaR for two  portfolios of six FX
majors was observed to be significant.

For a comprehensive modeling of multivariate dependence in finance
or insurance,  there are other issues in data analysis that should
be addressed carefully, such as time-dependent correlation
parameters and validation. These are not considered in the present
study.

\section{Acknowledgement}
\label{sec:acknowledgementodel}

We would like to thank Gareth Peters and John Donnelly for helpful
discussions and comments on the manuscript.

%\bibliography{bibliography}
%\bibliographystyle{my_bibstyle}

\renewcommand{\baselinestretch}{1.2}
\pagebreak

\begin{figure}
\centerline{\includegraphics[scale=0.5]{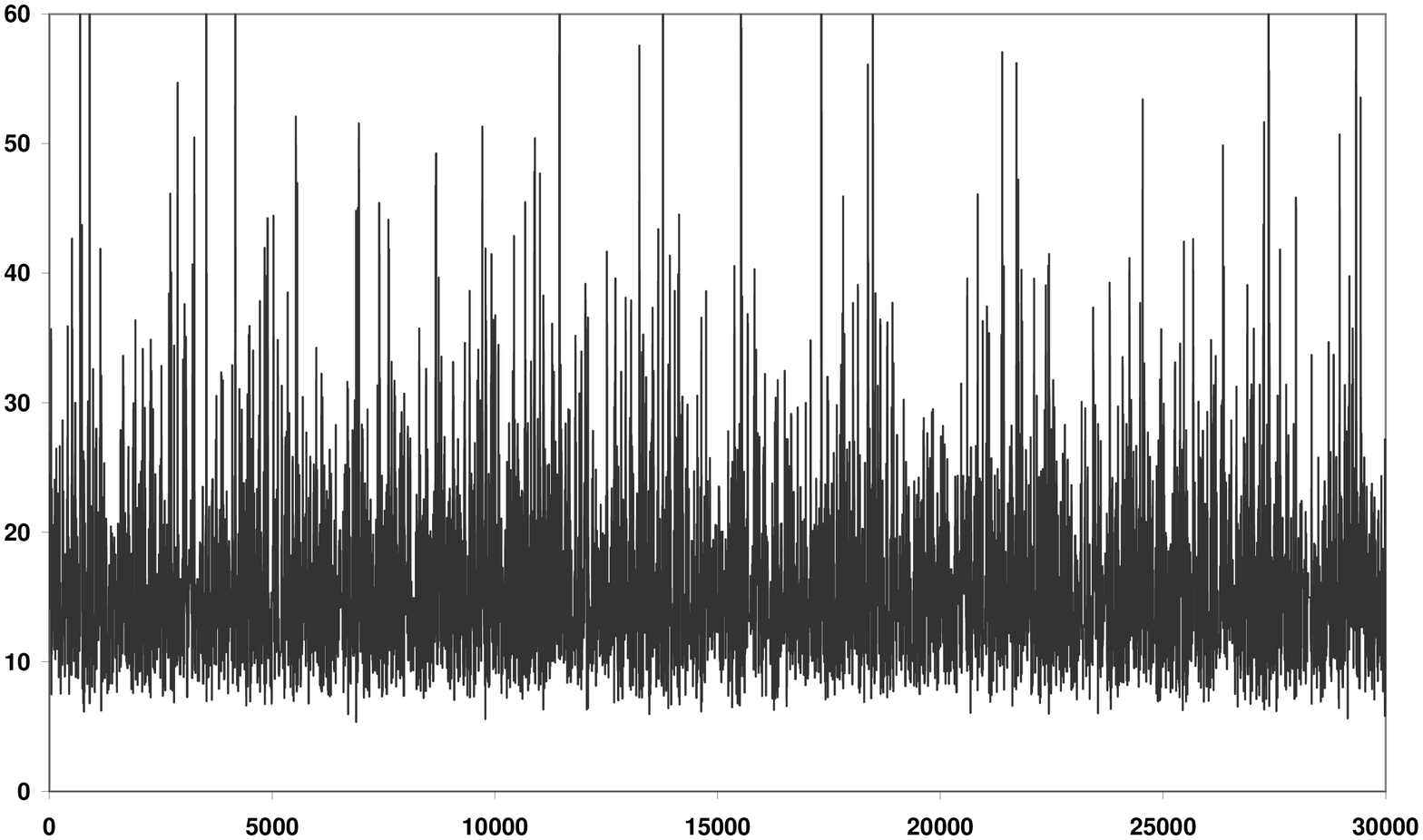}} \caption{Markov
chain paths for parameter $\nu _1^{(0)} $ of model $M_0 $.}
\label{fig1}
\end{figure}

\begin{figure}[htbp]
\centerline{\includegraphics[scale=0.5]{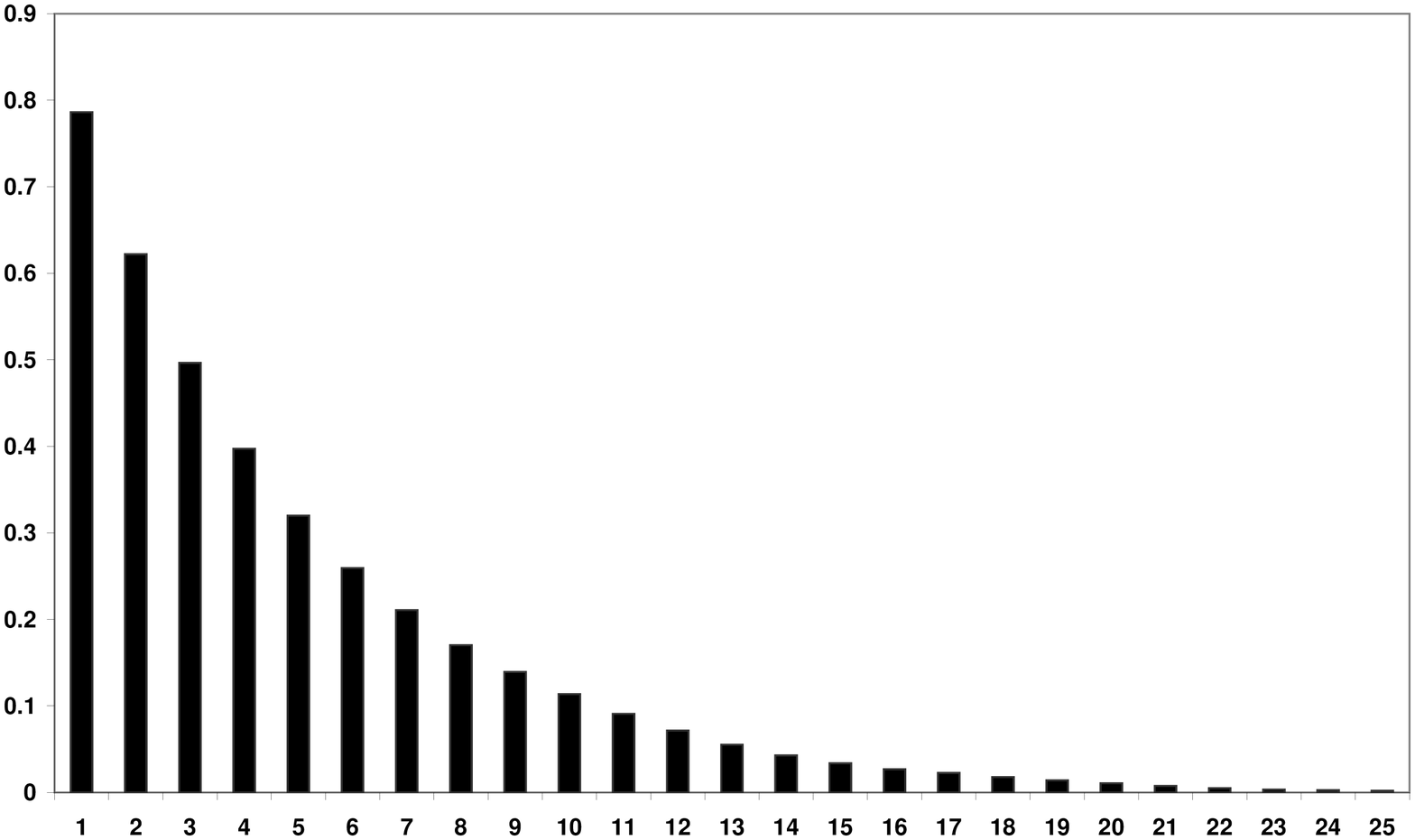}}
\caption{Autocorrelation of Markov chain samples for dof parameter
$\nu _1^{(0)}$  of model $M_0 $.} \label{fig2}
\end{figure}

\begin{table}
\caption{Group configurations and parameters for the 33 t-copula
models.}
\begin{tabular*}{1.0\textwidth}{clc@{\hspace{0.2\textwidth}}clc@{\hspace{0.2\textwidth}}clc@{\hspace{0.2\textwidth}}clc@{\hspace{0.3\textwidth}}}
\toprule \textbf{Model}& \textbf{\hspace{0.1\textwidth} Group
Configuration}&
\textbf{Parameters} \\
\midrule
$\mbox{M}_0 $&  (AUD), (CAD), (CHF), (EUR), (GBP), (JPY)&
$\nu _1^{(0)} ,\;\nu _2^{(0)} ,\;\nu _3^{(0)} ,\;\nu _4^{(0)} ,\;\nu _5^{(0)} ,\nu _6^{(0)} $ \\
$\mbox{M}_1 $& (AUD, CAD, CHF), (EUR, GBP, JPY)&
$\nu _1^{(1)} ,\quad \nu _2^{(1)} $ \\
$\mbox{M}_2 $& (AUD, CAD, EUR), (CHF, GBP, JPY)&
$\nu _1^{(2)} ,\quad \nu _2^{(2)} $ \\
$\mbox{M}_3 $& (AUD, CAD, GBP), (CHF, EUR, JPY)&
$\nu _1^{(3)} ,\quad \nu _2^{(3)} $ \\
$\mbox{M}_4 $& (AUD, CAD, JPY), (CHF, EUR, GBP)&
$\nu _1^{(4)} ,\quad \nu _2^{(4)} $ \\
$\mbox{M}_5 $& (AUD, CHF, EUR), (CAD, GBP, JPY)&
$\nu _1^{(5)} ,\quad \nu _2^{(5)} $ \\
$\mbox{M}_6 $& (AUD, CHF, GBP), (CAD, EUR, JPY)&
$\nu _1^{(6)} ,\quad \nu _2^{(6)} $ \\
$\mbox{M}_7 $& (AUD, CHF, JPY), (CAD, EUR, GBP)&
$\nu _1^{(7)} ,\quad \nu _2^{(7)} $ \\
$\mbox{M}_8 $& (AUD, EUR, GBP), (CAD, CHF, JPY)&
$\nu _1^{(8)} ,\quad \nu _2^{(8)} $ \\
$\mbox{M}_9 $& (AUD, EUR, JPY), (CAD, CHF, GBP)&
$\nu _1^{(9)} ,\quad \nu _2^{(9)} $ \\
$\mbox{M}_{10}$& (AUD, GBP, JPY), (CAD, CHF, EUR)&
$\nu _1^{(10)} ,\quad \nu _2^{(10)} $ \\
$\mbox{M}_{11} $& (GBP, JPY), (AUD, CAD, CHF, EUR)&
$\nu _1^{(11)},  \quad \nu _2^{(11)}$ \\
$\mbox{M}_{12} $& (AUD, CAD), (CHF, EUR, GBP, JPY)&
$\nu _1^{(12)},  \quad \nu _2^{(12)}$ \\
$\mbox{M}_{13} $& (AUD, CHF), (CAD, EUR, GBP, JPY)&
$\nu _1^{(13)},  \quad \nu _2^{(13)}$ \\
$\mbox{M}_{14} $& (AUD, EUR), (CAD, CHF, GBP, JPY)&
$\nu _1^{(14)},  \quad \nu _2^{(14)}$ \\
$\mbox{M}_{15} $& (AUD, GBP), (CAD, CHF, EUR, JPY)&
$\nu _1^{(15)},  \quad \nu _2^{(15)}$ \\
$\mbox{M}_{16} $& (AUD, JPY), (CAD, CHF, EUR, GBP)&
$\nu _1^{(16)},  \quad \nu _2^{(16)}$ \\
$\mbox{M}_{17} $& (CAD, CHF), (AUD, EUR, GBP, JPY)&
$\nu _1^{(17)},  \quad \nu _2^{(17)}$ \\
$\mbox{M}_{18} $& (CAD, EUR), (AUD, CHF, GBP, JPY)&
$\nu _1^{(18)},  \quad \nu _2^{(18)}$ \\
$\mbox{M}_{19} $& (CAD, GBP), (AUD, CHF, EUR, JPY)&
$\nu _1^{(19)},  \quad \nu _2^{(19)}$ \\
$\mbox{M}_{20} $& (CAD, JPY), (AUD, CHF, EUR, GBP)& $\nu
_1^{(20)},  \quad \nu _2^{(20)}$ \\
$\mbox{M}_{21} $& (CHF, EUR), (AUD, CAD, GBP, JPY)&
$\nu _1^{(21)},  \quad \nu _2^{(21)}$ \\
$\mbox{M}_{22} $& (CHF, GBP), (AUD, CAD, EUR, JPY)&
$\nu _1^{(22)},  \quad \nu _2^{(22)}$ \\
$\mbox{M}_{23} $& (CHF, JPY), (AUD, CAD, GBP, EUR)&
$\nu _1^{(23)},  \quad \nu _2^{(23)}$ \\
$\mbox{M}_{24} $& (EUR, GBP), (AUD, CAD, CHF, JPY)&
$\nu _1^{(24)},  \quad \nu _2^{(24)}$ \\
$\mbox{M}_{25} $& (EUR, JPY), (AUD, CAD, GBP, CHF)&
$\nu _1^{(25)},  \quad \nu _2^{(25)}$ \\
$\mbox{M}_{26} $& (AUD), (CAD, CHF, EUR, GBP, JPY)&
$\nu _1^{(26)},  \quad \nu _2^{(26)}$ \\
$\mbox{M}_{27} $& (CAD), (AUD, CHF, EUR, GBP, JPY)&
$\nu _1^{(27)},  \quad \nu _2^{(27)}$ \\
$\mbox{M}_{28} $& (CHF), (CAD, AUD, EUR, GBP, JPY)&
$\nu _1^{(28)},  \quad \nu _2^{(28)}$ \\
$\mbox{M}_{29} $& (EUR), (CAD, CHF, AUD, GBP, JPY)&
$\nu _1^{(29)},  \quad \nu _2^{(29)}$ \\
$\mbox{M}_{30} $& (GBP), (CAD, CHF, EUR, AUD, JPY)&
$\nu _1^{(30)},  \quad \nu _2^{(30)}$ \\
$\mbox{M}_{31} $& (JPY), (CAD, CHF, EUR, GBP, AUD)&
$\nu _1^{(31)},  \quad \nu _2^{(31)}$ \\
$\mbox{M}_{32} $& (AUD, CAD, CHF, EUR, GBP, JPY)&
$\nu _1^{(32)} $ \\

\bottomrule
\end{tabular*}
 \label{tab1}
\end{table}

\begin{table}[htbp]
\begin{center}
\caption{Autocorrelation estimates and corresponding cut-off lag
number.}
\begin{tabular*}
{0.6\textwidth}{clc@{\hspace{0.05\textwidth}}clc@{\hspace{0.05\textwidth}}clc@{\hspace{0.05\textwidth}}clc@{\hspace{0.05\textwidth}}clc@{\hspace{0.05\textwidth}}clc@{\hspace{0.05\textwidth}}clc@{\hspace{0.1\textwidth}}}
\toprule Parameter& $\nu _1^{(0)}$ & $\nu _2^{(0)}$ & $\nu _3^{(0)}$
& $\nu _4^{(0)} $ & $\nu _5^{(0)} $ & $\nu _6^{(0)} $ \\
 \midrule
$\hat {\tau }_k $ & 8.79 & 2.23 & 23.5 & 23.1 & 8.41 & 8.69  \\
$g_k^{\max } $  & 23 & 14 & 65 & 57 & 30 & 34 \\
\bottomrule
\end{tabular*}
\label{tab2}
\end{center}
\end{table}

\begin{table}[htbp]
\begin{center}
\caption{MCMC output values of posterior mean, standard error and
log-likelihood.}
\begin{tabular*}
{1.0\textwidth}{clc@{\hspace{0.2\textwidth}}clc@{\hspace{0.2\textwidth}}clc@{\hspace{0.2\textwidth}}clc@{\hspace{0.3\textwidth}}}
\toprule \textbf{Model}& \textbf{Posterior Mean (Standard Error)}&
\textbf{Log-likelihood} \\
\midrule \multirow{2}{*}{$M_0$ } &  {$\nu _1^{(0)} = 15.4(0.79)$,
\quad $\nu _2^{(0)} = 67.3(1.3)$, \quad $\nu
_3^{(0)} = 8.76(0.34)$,  } & \multirow{2}{*}{2353.1} \\
&{$\nu _4^{(0)}= 6.38(0.20)$, \quad $\nu _5^{(0)} =
11.6(0.46)$,\quad $\nu
_6^{(0)} = 18.3(1.1)$}&\\
$M_1 $& $\nu _1^{(1)} = 15.1\;(0.29),\quad \nu _2^{(1)} =
9.37\;(0.14)$&
2342.8 \\
$M_2 $& $\nu _1^{(2)} = 10.2\;(0.15),\quad \nu _2^{(2)} =
13.4\;(0.21)$&
2338.6 \\
$M_3 $& $\nu _1^{(3)} = 18.2\;(0.28),\quad \nu _2^{(3)} =
8.56\;(0.09)$&
2341.9 \\
$M_4 $& $\nu _1^{(4)} = 24.4\;(0.82),\quad \nu _2^{(4)} =
7.78\;(0.16)$&
2343.8 \\
$M_5 $& $\nu _1^{(5)} = 8.51\;(0.13),\quad \nu _2^{(5)} =
18.7\;(0.49)$&
2341.9 \\
$M_6 $& $\nu _1^{(6)} = 13.4\;(0.26),\quad \nu _2^{(6)} =
10.2\;(0.16)$&
2338.6 \\
$M_7 $& $\nu _1^{(7)} = 13.6\;(0.27),\quad \nu _2^{(7)} =
10.3\;(0.17)$&
2338.6 \\
$M_8 $& $\nu _1^{(8)} = 9.24(0.13),\quad \nu _2^{(8)} =
15.7\;(0.35)$&
2343.2 \\
$M_9 $& $\nu _1^{(9)} = 8.76(0.13),\quad \nu _2^{(9)} =
14.0\;(0.32)$&
2343.2 \\
$M_{10} $& $\nu _1^{(10)} = 12.6(0.23),\quad \nu _2^{(10)} =
11.1\;(0.17)$&
2336.7 \\
$M_{11} $& $\nu _1^{(11)} = 27.9(4.87),\quad \nu _2^{(11)} =
8.6\;(0.09)$&
2336.7 \\
$M_{12} $& $\nu _1^{(12)} = 14.1(0.44),\quad \nu _2^{(12)} =
10.6\;(0.14)$&
2336.7 \\
$M_{13} $& $\nu _1^{(13)} = 8.56(0.11),\quad \nu _2^{(13)} =
13.8\;(0.25)$&
2343.6 \\
$M_{14} $& $\nu _1^{(14)} = 13.2(0.74),\quad \nu _2^{(14)} =
11.1\;(0.14)$&
2336.7 \\
$M_{15} $& $\nu _1^{(15)} = 13.3(0.94),\quad \nu _2^{(15)} =
11.2\;(0.17)$&
2336.6 \\
$M_{16} $& $\nu _1^{(16)} = 17.4(0.85),\quad \nu _2^{(16)} =
9.78\;(0.13)$&
2343.0 \\
$M_{17} $& $\nu _1^{(17)} = 9.78(0.14),\quad \nu _2^{(17)} =
12.9\;(0.2)$&
2338.9 \\
$M_{18} $& $\nu _1^{(18)} = 24.9(4.4),\quad \nu _2^{(18)} =
8.96\;(0.09)$&
2342.5 \\
$M_{19} $& $\nu _1^{(19)} = 32.5(7.99),\quad \nu _2^{(19)} =
8.76\;(0.08)$&
2343.4 \\
$M_{20} $& $\nu _1^{(20)} = 7.46(0.11),\quad \nu _2^{(20)} =
16.8\;(0.59)$&
2343.0 \\
$M_{21} $& $\nu _1^{(21)} = 14.2(0.43),\quad \nu _2^{(21)} =
10.5\;(0.14)$&
2338.6 \\
$M_{22} $& $\nu _1^{(22)} = 14.3(0.49),\quad \nu _2^{(22)} =
10.5\;(0.15)$&
2338.4 \\
$M_{23} $& $\nu _1^{(23)} = 8.64(0.11),\quad \nu _2^{(23)} =
14.2\;(0.32)$&
2343.7 \\
$M_{24} $& $\nu _1^{(24)} = 8.66(0.09),\quad \nu _2^{(24)} =
13.6\;(0.23)$&
2343.4 \\
$M_{25} $& $\nu _1^{(25)} = 12.8(0.50),\quad \nu _2^{(25)} =
11.2\;(0.13)$&
2336.7 \\
$M_{26} $& $\nu _1^{(26)} = 15.2(1.63),\quad \nu _2^{(26)} =
11.3\;(0.31)$&
2336.4 \\
$M_{27} $& $\nu _1^{(27)} = 64.7(3.01),\quad \nu _2^{(27)} =
9.26\;(0.24)$&
2346.7 \\
$M_{28} $& $\nu _1^{(28)} = 16.0(1.15),\quad \nu _2^{(28)} =
11.0\;(0.38)$&
2338.4 \\
$M_{29} $& $\nu _1^{(29)} = 7.94(0.32),\quad \nu _2^{(29)} =
12.9\;(0.48)$&
2344.4 \\
$M_{30} $& $\nu _1^{(30)} = 15.1(1.53),\quad \nu _2^{(30)} =
11.3\;(0.32)$&
2336.5 \\
$M_{31} $& $\nu _1^{(31)} = 11.4(0.32),\quad \nu _2^{(31)} =
1.81\;(0.17)$&
2336.4 \\
$M_{32} $& $\nu _1^{(32)} = 11.4\;(0.14)$& 2336.8 \\
\bottomrule
\end{tabular*}
 \label{tab3}
\end{center}
\end{table}

\begin{table}[htbp]
\caption{MCMC output values of posterior mode, corresponding
log-likelihood, likelihood ratio $\Lambda$ and p-value comparing
$M_h$ with $M_0$.}
\begin{tabular*}
{1.03\textwidth}{clc@{\hspace{0.045\textwidth}}clc@{\hspace{0.1\textwidth}}clc@{\hspace{0.01\textwidth}}clc@{\hspace{0.1\textwidth}}clc@{\hspace{0.1\textwidth}}}
\toprule Model& MCMC posterior mode& Log-likelihood& $ - 2\log
(\Lambda )$&
p-value  \\
\toprule $M_0 $& ${\bm \nu }^{(0)} =
(11.5,82.4,7.92,5.81,10.3,14.3)$& 2354.3& 0&
N/A \\
 $M_1 $& ${\bm \nu }^{(1)} = (14.0,\quad 8.96)$& 2342.9& 22.8&
0.00014 \\
 $M_2 $& ${\bm \nu }^{(2)} = (9.75,\quad 12.6)$& 2338.7& 31.2&
$<$0.00001 \\
 $M_3 $& ${\bm \nu }^{(3)} = (16.6,\quad 8.22)$& 2342.1& 24.4&
$<$0.0001 \\
 $M_4 $& ${\bm \nu }^{(4)} = (21.0,\quad 7.49)$& 2344.1& 20.4&
0.00042 \\
 $M_5 $& ${\bm \nu }^{(5)} = (8.17,\quad 16.9)$& 2342.1&
24.4&
$<$0.0001 \\
 $M_6 $& ${\bm \nu }^{(6)} = (12.6,\quad 9.73)$& 2338.8& 31.0&
$<$0.00001 \\
 $M_7 $& ${\bm \nu }^{(7)} = (12.8,\quad 9.79)$& 2338.7&
31.2&
$<$0.00001 \\
 $M_8 $& ${\bm \nu }^{(8)} = (8.84,\quad 14.4)$& 2343.4& 21.8&
0.00022 \\
 $M_9 $& ${\bm \nu }^{(9)} = (8.76,\quad 14.0)$& 2343.3& 22.0&
0.00020 \\
 $M_{10} $& ${\bm \nu }^{(10)} = (11.8,\quad 10.5)$& 2336.9& 34.8&
$<$0.000001 \\
 $M_{11} $& ${\bm \nu }^{(11)} = (22.6,\quad 8.68)$& 2343.0 & 22.6&
0.00015 \\
 $M_{12} $& ${\bm \nu }^{(12)} = (13.1,\quad 10.2)$& 2338.6 & 31.5&
$<$0.00001 \\
 $M_{13} $& ${\bm \nu }^{(13)} = (8.23,\quad 13.1)$& 2343.7 & 21.3&
0.00028 \\
 $M_{14} $& ${\bm \nu }^{(14)} = (11.9,\quad 10.7)$& 2336.9 & 34.8&
$<$0.000001 \\
$M_{15} $& ${\bm \nu }^{(15)} = (11.7,\quad 10.8)$& 2336.8 & 34.9&
$<$0.000001 \\
 $M_{16} $& ${\bm \nu }^{(16)} = (15.7,\quad 9.39)$& 2343.2 & 22.2&
0.00018 \\
 $M_{17} $& ${\bm \nu }^{(17)} = (9.31,\quad 12.7)$& 2339.0 & 30.6&
$<$0.00001 \\
 $M_{18} $& ${\bm \nu }^{(18)} = (20.9,\quad 8.66)$& 2342.7 & 23.1&
0.00012 \\
$M_{19} $& ${\bm \nu }^{(19)} = (25.2,\quad 8.48)$& 2343.8 & 21.1&
0.00030 \\
 $M_{20} $& ${\bm \nu }^{(20)} = (7.14,\quad 15.7)$& 2343.1 & 22.3&
0.00017 \\
$M_{21} $& ${\bm \nu }^{(21)} = (13.2,\quad 10.1)$& 2338.5 & 31.5&
$<$0.00001 \\
 $M_{22} $& ${\bm \nu }^{(22)} = (13.3,\quad 10.1)$& 2338.5 & 31.5&
$<$0.00001 \\
 $M_{23} $& ${\bm \nu }^{(23)} = (8.27,\quad 13.4)$& 2343.8 & 21.0&
0.00031 \\
 $M_{24} $& ${\bm \nu }^{(24)} = (8.31,\quad 12.9)$& 2343.5 & 21.7&
0.00023 \\
 $M_{25} $& ${\bm \nu }^{(25)} = (11.7,\quad 10.8)$& 2336.8 & 34.9&
$<$0.000001 \\
$M_{26} $& ${\bm \nu }^{(26)} = (11.8,\quad 10.9)$& 2336.8 & 34.9&
$<$0.000001 \\
 $M_{27} $& ${\bm \nu }^{(27)} = (68.3,\quad 9.03)$& 2346.8 & 15.1&
0.0045 \\
 $M_{28} $& ${\bm \nu }^{(28)} = (14.0,\quad 10.5)$& 2338.6 & 31.4&
$<$0.00001 \\
 $M_{29} $& ${\bm \nu }^{(29)} = (7.55,\quad 12.2)$& 2344.5 & 19.6&
0.00060 \\
$M_{30} $& ${\bm \nu }^{(30)} = (12.0,\quad 10.9)$& 2336.8 & 34.9&
$<$0.000001 \\
 $M_{31} $& ${\bm \nu }^{(31)} = (11.4,\quad 11.0)$& 2336.8 & 35.0&
$<$0.000001 \\
$M_{32} $& ${\bm \nu }^{(32)} = (11.1)$& 2336.9& 34.9&
$<$0.00001 \\
\bottomrule
\end{tabular*}
\label{tab4}
\end{table}

\begin{table}[htbp]
\begin{center}
\caption{Bayes factors $B_{0i}$, DIC and model probabilities of all
candidates.}
\begin{tabular*}
{0.9\textwidth}{clc@{\hspace{0.1\textwidth}}clc@{\hspace{0.1\textwidth}}clc@{\hspace{0.1\textwidth}}clc@{\hspace{0.1\textwidth}}clc@{\hspace{0.1\textwidth}
clc@{\hspace{0.1\textwidth}}clc@{\hspace{0.1\textwidth}}clc@{\hspace{0.06\textwidth}}clc@{\hspace{0.1\textwidth}}clc@{\hspace{0.1\textwidth}}}}
\toprule Model& Log($B_{0h}$) & DIC &  Model prob. $(\%)$ &  Model prob. $(\%)$ excl. $M_0 $\\
\midrule
$M_0 $ & 0 & 0 &  88.5 & \hspace{0.1\textwidth} N/A \\
$M_1 $ & 16.7 & 15.8 &  0.15 & \hspace{0.1\textwidth} 1.28 \\
$M_2 $ & 20.1 & 24.0 &  $<0.1 $ &\hspace{0.1\textwidth} $<0.1 $\\
$M_3 $ & 16.6 & 17.5 &  $<0.1 $ &\hspace{0.1\textwidth} 0.54 \\
$M_4 $ & 16.4 & 13.0 &  0.45    &\hspace{0.1\textwidth} 3.89 \\
$M_5 $ & 18.7 & 17.4 &  $<0.1 $ &\hspace{0.1\textwidth} 0.55 \\
$M_6 $ & 19.3 & 24.1 &  $<0.1 $ &\hspace{0.1\textwidth} $<0.1 $\\
$M_7 $ & 21.5 & 24.3 &  $<0.1 $ &\hspace{0.1\textwidth} $<0.1 $\\
$M_8 $ & 14.5 & 15.0 &  0.12    &\hspace{0.1\textwidth} 1.07 \\
$M_9 $ & 18.9 & 15.1 &  0.22    &\hspace{0.1\textwidth} 1.89 \\
$M_{10}$ & 21.9 & 27.8 & $<0.1$ &\hspace{0.1\textwidth} $<0.1$ \\
$M_{11}$ & 18.0 & 15.6 &  0.15  &\hspace{0.1\textwidth} 1.26 \\
$M_{12}$ & 24.9 & 24.7 & $<0.1$ &\hspace{0.1\textwidth} $<0.1$\\
$M_{13}$ & 14.9 & 14.3 &  0.31        &\hspace{0.1\textwidth} 2.68\\
$M_{14}$ & 23.2 & 27.9 &  $<0.1 $     &\hspace{0.1\textwidth}  $<0.1$\\
$M_{15}$ & 25.5 & 28.0 &  $<0.1 $     &\hspace{0.1\textwidth}  $<0.1$\\
$M_{16} $ & 14.2 & 15.3 &  0.20       &\hspace{0.1\textwidth}  1.73\\
$M_{17} $ & 21.7 & 23.7 &  $<0.1 $    &\hspace{0.1\textwidth}  $<0.1$\\
$M_{18} $ & 15.3 & 16.1  &  0.12       &\hspace{0.1\textwidth}  1.03\\
$M_{19} $ & 15.1 & 14.1 &  0.29       &\hspace{0.1\textwidth}  2.52 \\
$M_{20} $ & 15.1 & 15.3 &  0.19       &\hspace{0.1\textwidth}  1.64\\
$M_{21} $ & 21.5 & 24.3 &  $<0.1 $    &\hspace{0.1\textwidth}  $<0.1$\\
$M_{22} $ & 20.6 & 24.6 &  $<0.1 $    &\hspace{0.1\textwidth}  $<0.1$\\
$M_{23} $ & 17.1 & 14.1 &  0.36       &\hspace{0.1\textwidth}  3.10\\
$M_{24} $ & 15.4 & 14.7 &  0.25       &\hspace{0.1\textwidth}  2.22 \\
$M_{25} $ & 22.6 & 27.9 &  $<0.1 $    &\hspace{0.1\textwidth}  $<0.1$\\
$M_{26} $ & 22.2 & 28.0 & $<0.1$      &\hspace{0.1\textwidth}  $<0.1$ \\
$M_{27} $ & 11.0 & 6.73&   7.8        &\hspace{0.1\textwidth}  68.3\\
$M_{28} $ & 21.0 & 24.6& $<0.1$       &\hspace{0.1\textwidth}  $<0.1$\\
$M_{29} $ & 16.3 & 12.7&  0.7         &\hspace{0.1\textwidth}  6.18\\
$M_{30} $ & 28.0 & 28.0& $<0.1$       &\hspace{0.1\textwidth} $<0.1$ \\
$M_{31} $ & 22.8 & 28.1& $<0.1$       &\hspace{0.1\textwidth} $<0.1$\\
$M_{32} $ & 22.3 & 28.1& $<0.1$       &\hspace{0.1\textwidth} $<0.1$ \\
\bottomrule
\end{tabular*}
\label{tab6}
\end{center}
\end{table}

\begin{table}[htbp]
\begin{center}
 \caption{The 0.99 conditional Value-at-Risk ($CVaR_{0.99}$)
predicted by model $M_0 $ and $M_{27} $, for two portfolios of six
major currencies. $\delta = (CVaR_{0.99}^{M_{27} } -
CVaR_{0.99}^{M_0 } ) / CVaR_{0.99}^{M_0 } $ is the relative
difference of CVaR between $M_{27} $ and $M_0 $. Standard errors are
in parentheses.}
\begin{tabular*}
{1.0\textwidth}{clc@{\hspace{0.1\textwidth}}clc@{\hspace{0.1\textwidth}}clc@{\hspace{0.1\textwidth}}clc@{\hspace{0.1\textwidth}}clc@{\hspace{0.1\textwidth}
clc@{\hspace{0.1\textwidth}}clc@{\hspace{0.1\textwidth}}clc@{\hspace{0.1\textwidth}}clc@{\hspace{0.1\textwidth}}}}
\toprule  {Portfolio asset weights $w_i $ } &
\multirow{2}{*}{$CVaR_{0.99}^{(M_0 )} $} &
\multirow{2}{*}{$CVaR_{0.99}^{(M_{27} )} $} & \multirow{2}{*}{$\delta $} \\
 {(AUD, CAD, CHF, EUR, GBP, JPY)} & \\
 \midrule
 (0.25, 0.25, 0.8, -0.8, 0.25, 0.25) & 1.707 (0.004)& 1.425 (0.003)&
-16.5{\%} \\
(0.25, 0.25, -0.8, 0.8, 0.25, 0.25)& 1.737 (0.003)& 1.782 (0.003)&
2.6{\%} \\
\bottomrule
\end{tabular*}
\label{tab7}
\end{center}
\end{table}

\begin{table}[htbp]
\begin{center}
 \caption{The 0.99 conditional Value-at-Risk  ($CVaR_{0.99}$)
predicted by model $M_0 $ and $M_4 $, for two portfolios of six
major currencies. $\delta = (CVaR_{0.99}^{M_4 } - CVaR_{0.99}^{M_0 }
) / CVaR_{0.99}^{M_0 } $ is the relative difference of CVaR between
$M_4 $ and $M_0 $. Standard errors are in parentheses.}
\begin{tabular*}
{1.0\textwidth}{clc@{\hspace{0.1\textwidth}}clc@{\hspace{0.1\textwidth}}clc@{\hspace{0.1\textwidth}}clc@{\hspace{0.1\textwidth}}clc@{\hspace{0.1\textwidth}
clc@{\hspace{0.1\textwidth}}clc@{\hspace{0.1\textwidth}}clc@{\hspace{0.1\textwidth}}clc@{\hspace{0.1\textwidth}}}}
\toprule  {Portfolio asset weights $w_i $ } &
\multirow{2}{*}{$CVaR_{0.99}^{(M_0 )} $} &
\multirow{2}{*}{$CVaR_{0.99}^{(M_4 )} $} & \multirow{2}{*}{$\delta $} \\
 {(AUD, CAD, CHF, EUR, GBP, JPY)} & \\
 \midrule
 (0.25, 0.25, 0.8, -0.8, 0.25, 0.25) & 1.571 (0.004)& 1.366 (0.003)&
-13.0{\%} \\
(0.25, 0.25, -0.8, 0.8, 0.25, 0.25)& 1.608 (0.003)& 1.732 (0.003)&
7.7{\%} \\
\bottomrule
\end{tabular*}
\label{tab8}
\end{center}
\end{table}

\end{document}